\documentclass[10pt,conference]{IEEEtran}

\usepackage{cite}
\usepackage{amsmath,amssymb,amsfonts}
\usepackage{graphicx}
\usepackage{textcomp}
\usepackage{xcolor}
\usepackage{booktabs}
\usepackage{multirow}
\usepackage{tabularx}
\usepackage[caption=false,font=footnotesize]{subfig}
\usepackage{url}

\newcolumntype{Y}{>{\raggedright\arraybackslash}X}
\newcolumntype{L}[1]{>{\raggedright\arraybackslash}p{#1}}

\def\BibTeX{{\rm B\kern-.05em{\sc i\kern-.025em b}\kern-.08em
    T\kern-.1667em\lower.7ex\hbox{E}\kern-.125emX}}

\newcommand{\findingbox}[2]{%
\par\smallskip\noindent
\fbox{\begin{minipage}{0.96\linewidth}
\footnotesize\textbf{#1.} #2
\end{minipage}}
\par\smallskip}

\begin{document}

\title{DSL-Bench: Evaluating and Explaining LLM Code Generation Across Domain-Specific Languages}

\author{
\IEEEauthorblockN{Zihan Zhou, Mingxi Ye, Yan Gao, Guipeng Wei, and Zibin Zheng}
\IEEEauthorblockA{Sun Yat-sen University}
}

\maketitle

\providecommand{\ModelScorecardTable}{}
\renewcommand{\ModelScorecardTable}{%
\begin{table*}[t]
\centering
\scriptsize
\setlength{\tabcolsep}{4pt}
\caption{LLM capability profiles with DSL-bootstrap uncertainty (\(B=2000\)).}
\label{tab:supp-model-scorecard}
\begin{tabular}{llrrr}
\toprule
Model & Score & Mean & CI low & CI high \\
\midrule
DeepSeek-v3.2 & Semantic & 18.41 & 6.72 & 34.40 \\
DeepSeek-v3.2 & Syntax & 47.65 & 24.11 & 70.11 \\
DeepSeek-v3.2 & K-Axis & 1.14 & 0.17 & 1.84 \\
DeepSeek-v3.2 & T-Axis & 1.08 & 0.01 & 1.91 \\
DeepSeek-v3.2 & S-Axis & 0.88 & 0.03 & 1.66 \\
DeepSeek-v3.2 & L-Axis & 1.05 & -0.41 & 2.08 \\
DeepSeek-v3.2 & Composite Axis & 1.04 & 0.26 & 1.68 \\
DeepSeek-v3.2 & Hard-DSL Robustness & 17.41 & 3.40 & 39.79 \\
GLM-4.7 & Semantic & 17.32 & 6.67 & 34.59 \\
GLM-4.7 & Syntax & 48.46 & 25.59 & 71.12 \\
GLM-4.7 & K-Axis & 1.07 & 0.55 & 1.47 \\
GLM-4.7 & T-Axis & 1.20 & 0.38 & 2.06 \\
GLM-4.7 & S-Axis & 1.52 & 0.76 & 2.22 \\
GLM-4.7 & L-Axis & 1.22 & -0.04 & 2.45 \\
GLM-4.7 & Composite Axis & 1.25 & 0.65 & 1.87 \\
GLM-4.7 & Hard-DSL Robustness & 16.32 & 3.59 & 39.96 \\
Qwen3-32B & Semantic & 12.82 & 4.27 & 24.72 \\
Qwen3-32B & Syntax & 35.97 & 15.80 & 59.17 \\
Qwen3-32B & K-Axis & 0.38 & -0.19 & 0.80 \\
Qwen3-32B & T-Axis & 0.16 & -0.36 & 0.65 \\
Qwen3-32B & S-Axis & -0.00 & -0.43 & 0.58 \\
Qwen3-32B & L-Axis & 0.09 & -0.41 & 0.46 \\
Qwen3-32B & Composite Axis & 0.16 & -0.19 & 0.48 \\
Qwen3-32B & Hard-DSL Robustness & 12.10 & 2.11 & 29.30 \\
Qwen3-8B & Semantic & 9.62 & 2.45 & 20.61 \\
Qwen3-8B & Syntax & 23.21 & 8.02 & 41.80 \\
Qwen3-8B & K-Axis & -0.03 & -0.66 & 0.89 \\
Qwen3-8B & T-Axis & -0.56 & -0.90 & 0.09 \\
Qwen3-8B & S-Axis & -0.23 & -0.72 & 0.69 \\
Qwen3-8B & L-Axis & -0.78 & -0.97 & -0.41 \\
Qwen3-8B & Composite Axis & -0.40 & -0.72 & 0.07 \\
Qwen3-8B & Hard-DSL Robustness & 9.18 & 1.06 & 25.18 \\
Qwen2.5-Coder-7B & Semantic & 9.43 & 2.70 & 20.30 \\
Qwen2.5-Coder-7B & Syntax & 27.87 & 11.01 & 48.68 \\
Qwen2.5-Coder-7B & K-Axis & -0.70 & -1.00 & -0.06 \\
Qwen2.5-Coder-7B & T-Axis & -0.41 & -0.85 & 0.35 \\
Qwen2.5-Coder-7B & S-Axis & -0.53 & -0.88 & -0.01 \\
Qwen2.5-Coder-7B & L-Axis & -0.18 & -0.75 & 0.89 \\
Qwen2.5-Coder-7B & Composite Axis & -0.45 & -0.77 & -0.01 \\
Qwen2.5-Coder-7B & Hard-DSL Robustness & 8.72 & 1.18 & 24.25 \\
Ministral-3-8B-Instruct & Semantic & 8.89 & 2.09 & 21.04 \\
Ministral-3-8B-Instruct & Syntax & 24.74 & 9.50 & 45.56 \\
Ministral-3-8B-Instruct & K-Axis & -0.75 & -1.16 & -0.15 \\
Ministral-3-8B-Instruct & T-Axis & -0.47 & -0.80 & -0.12 \\
Ministral-3-8B-Instruct & S-Axis & -0.74 & -1.13 & -0.40 \\
Ministral-3-8B-Instruct & L-Axis & -0.59 & -0.97 & -0.25 \\
Ministral-3-8B-Instruct & Composite Axis & -0.64 & -0.89 & -0.35 \\
Ministral-3-8B-Instruct & Hard-DSL Robustness & 8.23 & 0.98 & 25.93 \\
Qwen3-1.7B & Semantic & 4.96 & 1.23 & 11.24 \\
Qwen3-1.7B & Syntax & 11.72 & 3.32 & 26.97 \\
Qwen3-1.7B & K-Axis & -1.11 & -1.53 & -0.76 \\
Qwen3-1.7B & T-Axis & -0.99 & -1.42 & -0.58 \\
Qwen3-1.7B & S-Axis & -0.88 & -1.29 & -0.55 \\
Qwen3-1.7B & L-Axis & -0.81 & -1.10 & -0.41 \\
Qwen3-1.7B & Composite Axis & -0.95 & -1.21 & -0.67 \\
Qwen3-1.7B & Hard-DSL Robustness & 4.65 & 0.48 & 13.39 \\
\bottomrule
\end{tabular}
\end{table*}
}

\providecommand{\ScorecardTiesTable}{}
\renewcommand{\ScorecardTiesTable}{%
\begin{table*}[t]
\centering
\scriptsize
\setlength{\tabcolsep}{4pt}
\caption{Scorecard leaders and models tied by overlapping 95\% bootstrap CIs.}
\label{tab:supp-scorecard-ties}
\begin{tabularx}{\textwidth}{llrY}
\toprule
Score & Leader & Leader mean & Tied with leader by CI overlap \\
\midrule
Semantic & DeepSeek-v3.2 & 18.41 & DeepSeek-v3.2; GLM-4.7; Qwen3-32B; Qwen3-8B; Qwen2.5-Coder-7B; Ministral-3-8B-Instruct; Qwen3-1.7B \\
Syntax & GLM-4.7 & 48.46 & DeepSeek-v3.2; GLM-4.7; Qwen3-32B; Qwen3-8B; Qwen2.5-Coder-7B; Ministral-3-8B-Instruct; Qwen3-1.7B \\
K-Axis & DeepSeek-v3.2 & 1.14 & DeepSeek-v3.2; GLM-4.7; Qwen3-32B; Qwen3-8B \\
T-Axis & GLM-4.7 & 1.20 & DeepSeek-v3.2; GLM-4.7; Qwen3-32B \\
S-Axis & GLM-4.7 & 1.52 & DeepSeek-v3.2; GLM-4.7 \\
L-Axis & GLM-4.7 & 1.22 & DeepSeek-v3.2; GLM-4.7; Qwen3-32B; Qwen2.5-Coder-7B \\
Composite Axis & GLM-4.7 & 1.25 & DeepSeek-v3.2; GLM-4.7 \\
Hard-DSL Robustness & DeepSeek-v3.2 & 17.41 & DeepSeek-v3.2; GLM-4.7; Qwen3-32B; Qwen3-8B; Qwen2.5-Coder-7B; Ministral-3-8B-Instruct; Qwen3-1.7B \\
\bottomrule
\end{tabularx}
\end{table*}
}

\providecommand{\BaselineModelTable}{}
\renewcommand{\BaselineModelTable}{%
\begin{table}[t]
\centering
\scriptsize
\caption{Baseline model performance, macro-averaged across DSLs.}
\label{tab:supp-baseline-models}
\begin{tabular}{lrr}
\toprule
Model & Syntax (\%) & Semantic (\%) \\
\midrule
DeepSeek-v3.2 & 47.65 & 18.41 \\
GLM-4.7 & 48.46 & 17.32 \\
Qwen3-32B & 35.97 & 12.82 \\
Qwen3-8B & 23.21 & 9.62 \\
Qwen2.5-Coder-7B & 27.87 & 9.43 \\
Ministral-3-8B-Instruct & 24.74 & 8.89 \\
Qwen3-1.7B & 11.72 & 4.96 \\
\midrule
Mean & 31.37 & 11.64 \\
\bottomrule
\end{tabular}
\end{table}
}

\providecommand{\DSLBottleneckProfileTable}{}
\renewcommand{\DSLBottleneckProfileTable}{%
\begin{table*}[t]
\centering
\scriptsize
\setlength{\tabcolsep}{3pt}
\caption{DSL-Profiler bottleneck profiles. Difficulty=\(100-\)MeanSem; Risk averages KD, HC, inverse SF, and recency.}
\label{tab:supp-dsl-bottleneck-profile}
\begin{tabular}{lrrrrrl}
\toprule
DSL & MeanSem & Difficulty & SemGap & Disagree & Risk & Bottleneck \\
\midrule
Nextflow & 0.00 & 100.00 & 51.46 & 0.00 & 68.69 & domain-knowledge \\
UFL & 0.00 & 100.00 & 0.00 & 0.00 & 78.32 & domain-knowledge \\
OpenQASM 3 & 0.00 & 100.00 & 0.14 & 0.00 & 69.46 & version-interference \\
SysML v2 & 0.16 & 99.84 & 2.06 & 0.39 & 66.93 & legacy-version \\
Behavior Tree XML & 0.73 & 99.27 & 20.31 & 1.29 & 32.25 & structural-semantics \\
Lean & 0.75 & 99.25 & 0.00 & 1.22 & 54.58 & toolchain-harness \\
SMILES & 2.88 & 97.12 & 47.50 & 3.73 & 37.75 & mixed \\
Vega-Lite & 2.89 & 97.11 & 47.04 & 1.89 & 44.99 & domain-knowledge \\
jq & 5.04 & 94.96 & 8.50 & 2.66 & 45.16 & syntax/toolchain \\
Verilog-A & 5.16 & 94.84 & 5.56 & 6.87 & 66.74 & domain-knowledge \\
Terraform HCL & 8.08 & 91.92 & 6.74 & 6.49 & 47.34 & configuration-policy \\
Cypher & 13.79 & 86.21 & 67.42 & 6.58 & 57.45 & domain-knowledge \\
Solidity & 16.63 & 83.37 & 5.41 & 4.60 & 58.41 & domain-knowledge \\
IEC 61131-3 Structured Text & 19.82 & 80.18 & 17.05 & 10.41 & 37.24 & mixed \\
Dafny & 21.69 & 78.31 & 4.65 & 15.13 & 51.74 & syntax/toolchain \\
LaTeX & 22.31 & 77.69 & 33.73 & 9.24 & 34.46 & mixed \\
Verilog & 42.40 & 57.60 & 24.82 & 18.16 & 33.27 & mixed \\
GDScript & 47.14 & 52.86 & 12.85 & 11.19 & 37.70 & mixed \\
\bottomrule
\end{tabular}
\end{table*}
}

\providecommand{\AxisMembershipTable}{}
\renewcommand{\AxisMembershipTable}{%
\begin{table}[t]
\centering
\scriptsize
\caption{Scorecard bottleneck-axis membership.}
\label{tab:supp-axis-membership}
\begin{tabular}{ll}
\toprule
DSL & Axes \\
\midrule
Behavior Tree XML & S \\
Cypher & K,S \\
Dafny & K,S \\
GDScript & none \\
IEC 61131-3 Structured Text & T,L \\
LaTeX & none \\
Lean & T,S \\
Nextflow & K,T,S \\
OpenQASM 3 & K,L \\
SMILES & K \\
Solidity & K,T \\
SysML v2 & T,S,L \\
Terraform HCL & T,S,L \\
UFL & K,T \\
Vega-Lite & K,S \\
Verilog & T \\
Verilog-A & K,T,L \\
jq & none \\
\bottomrule
\end{tabular}
\end{table}
}

\providecommand{\ProfilerValidationTable}{}
\renewcommand{\ProfilerValidationTable}{%
\begin{table}[t]
\centering
\scriptsize
\caption{LOOCV DSL-Profiler validation with standardized factors and model fixed effects.}
\label{tab:supp-profiler-validation}
\begin{tabular}{lrr}
\toprule
Target & MAE & RMSE \\
\midrule
semantic & 11.78 & 14.60 \\
syntax & 20.29 & 24.77 \\
\bottomrule
\end{tabular}
\end{table}
}

\providecommand{\ProfilerCoverageTable}{}
\renewcommand{\ProfilerCoverageTable}{%
\begin{table}[t]
\centering
\scriptsize
\caption{DSL-Profiler semantic risk-interval coverage (LOOCV).}
\label{tab:supp-profiler-coverage}
\begin{tabular}{lrrr}
\toprule
Target & Covered & Total & Coverage \\
\midrule
semantic & 11 & 18 & 61.1\% \\
\bottomrule
\end{tabular}
\end{table}
}

\providecommand{\DSLBenchOverviewTable}{}
\renewcommand{\DSLBenchOverviewTable}{%
\begin{table*}[t]
\centering
\scriptsize
\setlength{\tabcolsep}{2.2pt}
\caption{DSL-Bench coverage, validators, and construction sources.}
\label{tab:dslbench-overview}
\begin{tabularx}{\textwidth}{L{0.135\textwidth}L{0.105\textwidth}L{0.085\textwidth}L{0.155\textwidth}L{0.265\textwidth}L{0.125\textwidth}}
\toprule
Domain & DSL & Source tasks & Syntax & Semantic basis & Construction \\
\midrule
Bioinformatics & Nextflow DSL2 & 83 & Nextflow stub run & module/dataflow/control-structure checks & New benchmark \\
Cheminformatics & SMILES & 33,008 & RDKit parser & canonical molecular equivalence & Reused \\
Data Engineering & jq & 4,392 & jq parser & JSON output equivalence & Reused \\
Data Visualization & Vega-Lite & 7,247 & JSON + schema & chart-intent structural checks & Reused \\
Database/Query & Cypher & 10,882 & Neo4j execution & normalized query-result equivalence & Reused \\
DevOps/Infra. & Terraform HCL & 458 & Terraform validate/plan & task-specific policy checks & Reused \\
Document Processing & LaTeX & 440 & \texttt{pdflatex} & document-structure intent checks & Reused \\
Game Development & GDScript & 30 & Godot preflight & runtime task harness & New benchmark \\
Hardware Design & Verilog & 156 & Icarus Verilog & testbench simulation & Reused \\
Industrial Automation & IEC ST & 31 & RuSTy-compatible check & interface/property checks & Validator rebuilt \\
Model-Driven Eng. & SysML v2 & 90 & SysML v2 validator & dependency and semantic diagnostics & Validator rebuilt \\
Power Electronics & Verilog-A & 36 & OpenVAF & ngspice behavioral tests & New benchmark \\
Program Verification & Dafny & 166 & Dafny resolve & Dafny verification & Reused \\
Quantum Computing & OpenQASM 3 & 397 & QASM parser/verifier & circuit-level checks & Reused \\
Robotics & BT XML & 588 & BehaviorTree.CPP & tree-structure/dataflow checks & Reused \\
Scientific Computing & UFL & 25 & Python/UFL preflight & symbolic form-signature equivalence & New benchmark \\
Smart Contract & Solidity & 1,125 & solc/Foundry & repository tests & Reused \\
Theorem Proving & Lean & 1,159 & Lean/lake & proof checking without \texttt{sorry}/\texttt{admit} & Reused \\
\midrule
Total & 18 DSLs & 60,313 & N/A & N/A & N/A \\
\bottomrule
\end{tabularx}
\end{table*}
}

\providecommand{\FactorSummaryTable}{}
\renewcommand{\FactorSummaryTable}{%
\begin{table}[t]
\centering
\scriptsize
\setlength{\tabcolsep}{2pt}
\caption{DSL-level factor trends against baseline semantic pass rate.}
\label{tab:factor-correlations}
\begin{tabularx}{\linewidth}{L{0.25\linewidth}rrrX}
\toprule
Factor & Slope & \(r\) & \(p\) & Interpretation \\
\midrule
Knowledge Demand & -2.31 & -0.508 & .030 & Largest observed negative association. \\
Harness Complexity & -0.83 & -0.150 & .560 & Weak semantic association; stronger for syntax. \\
Syntax Familiarity & +0.22 & +0.034 & .889 & Little guidance about semantic success. \\
Release Year & -4.49 & -0.351 & .144 & Newer DSLs trend lower, with high uncertainty. \\
\bottomrule
\end{tabularx}
\end{table}
}

\begin{abstract}
Large language models (LLMs) are increasingly used in software-engineering workflows, making reliable evaluation of LLM code generation important. Existing benchmarks largely focus on general-purpose languages, yet software systems in a wide range of real-world application domains rely on domain-specific languages (DSLs) with their own syntax, semantics, and validation requirements. Benchmarks designed for individual DSLs provide useful domain-specific evidence, but their designs and metrics are difficult to compare across DSLs.
We present DSL-Bench, an executable benchmark and analysis framework for cross-DSL code generation. Starting from 13,045 DSL-related papers, LLM-assisted open card sorting yields 18 representative application domains. For each domain, we select an evaluable DSL and instantiate a corresponding syntax checker and task-level semantic oracle. Evaluating 7 LLMs on 18 DSLs, we observe a mean syntax pass rate of 31.37\% and a mean semantic pass rate of only 11.64\%, demonstrating a large gap between syntax acceptance and semantic correctness. A variance decomposition shows that differences among DSLs account for 76.99\% of the variation in semantic pass rate, whereas differences among models account for 7.86\%. To characterize this variation, we analyze four DSL-level factors. Knowledge Demand shows the largest negative association with semantic correctness ($r$=-0.508). To assess intervention strategies, we test prompt engineering across all 18 DSLs and additionally evaluate five methods on five targeted DSLs. Prompt engineering yields uneven gains. In the five targeted DSLs, the strongest targeted method, RAG-Refinement, improves semantic correctness by 16.29 percentage points on average. We further derive diagnostic capability profiles and DSL-Profiler summaries for interpreting data-scarce DSLs. These results show that cross-DSL code generation is limited by toolchain-specific semantic bottlenecks and should be evaluated beyond syntax-level acceptance.
\end{abstract}


\section{Introduction}

Existing large language model (LLM) code generation benchmarks mainly evaluate general-purpose programming languages (GPLs). As LLMs are increasingly used to generate, repair, and assist with code, these benchmarks have become important for comparing models, exposing failure modes, and guiding coding agents. Benchmarks such as HumanEval, MBPP, APPS, MultiPL-E, DS-1000, and repository-level repair suites have shaped how the community measures LLM coding capability~\cite{chen2021codex,austin2021mbpp,hendrycks2021apps,li2022codecontests,lu2021codexglue,cassano2023multiple,lai2023ds1000,jimenez2024swebench}. Subsequent work has extended this evaluation landscape to broader domains, richer context, different code granularities, stronger tests, and contamination control~\cite{zhu2025domaineval,yu2024codereval,du2024classeval}.

Evaluating LLM performance on domain-specific language (DSL) code generation is also important, since a wide range of real-world software-engineering domains rely on DSLs. A DSL is a programming language tailored to a particular application domain, trading generality for domain-specific expressiveness~\cite{mernik2005dsl}. DSLs appear in hardware design, formal verification, infrastructure configuration, database and graph querying, workflow modeling, behavior modeling, and scientific computing. Examples include Verilog for hardware design~\cite{liu2023verilogeval}, Lean for theorem proving~\cite{yang2023leandojo}, and Cypher for graph queries~\cite{feng2025cypherbench}. A DSL is not simply a low-resource GPL or a GPL with a domain library. It encodes domain abstractions, toolchain constraints, and semantic rules directly into the target notation.

Evaluating DSL code generation is difficult because DSL correctness is tied to domain-specific semantics and toolchains.
For GPL benchmarks, generated programs are often evaluated through execution tests in relatively common programming environments, but test strength still affects the reliability of the conclusions~\cite{liu2023evalplus,jain2025livecodebench}.
For DSLs, the evaluator often needs a domain-specific checker or oracle.
Verilog generation depends on hardware simulation, Lean and Dafny depend on proof checking, infrastructure-as-code depends on resource and configuration constraints, and Cypher depends on graph-query results~\cite{liu2023verilogeval,yang2023leandojo,misu2024dafny,kon2024iaceval,feng2025cypherbench}.

Although a few benchmarks have been proposed for specific DSLs, cross-DSL LLM behavior remains unclear.
Prior work has studied hardware description~\cite{liu2023verilogeval}, theorem proving~\cite{yang2023leandojo,misu2024dafny}, infrastructure configuration~\cite{kon2024iaceval}, smart contracts~\cite{peng2025soleval}, graph query~\cite{feng2025cypherbench}, and other DSL settings.
These benchmarks are designed around their own tasks, oracles, metrics, and model sets.
As a result, they cannot easily show how LLM performance varies across DSLs, which DSL properties explain this variation, or which improvement strategies work under different DSL bottlenecks.

Therefore, evaluation results are needed from multiple DSLs under the same set of LLMs and testing conditions. Improvement methods also need to be tested across DSLs, so that their cross-DSL effects can be analyzed.

To meet this need, we present \textbf{DSL-Bench}, an executable benchmark and analysis framework for LLM code generation across DSLs.
DSL-Bench is constructed through a literature-based domain selection process and executable benchmark design.
Starting from 13,045 DSL-related papers, we identify representative application domains and select 18 evaluable DSLs.
For each DSL, we instantiate or adapt a syntax checker and a task-level semantic oracle, then evaluate the same set of LLMs under the same testing conditions.
DSL-Bench also analyzes DSL-level factors and tests improvement methods across DSLs, so that baseline performance and intervention effects can be studied together.

\begin{figure*}[t]
\centering
\includegraphics[width=\textwidth]{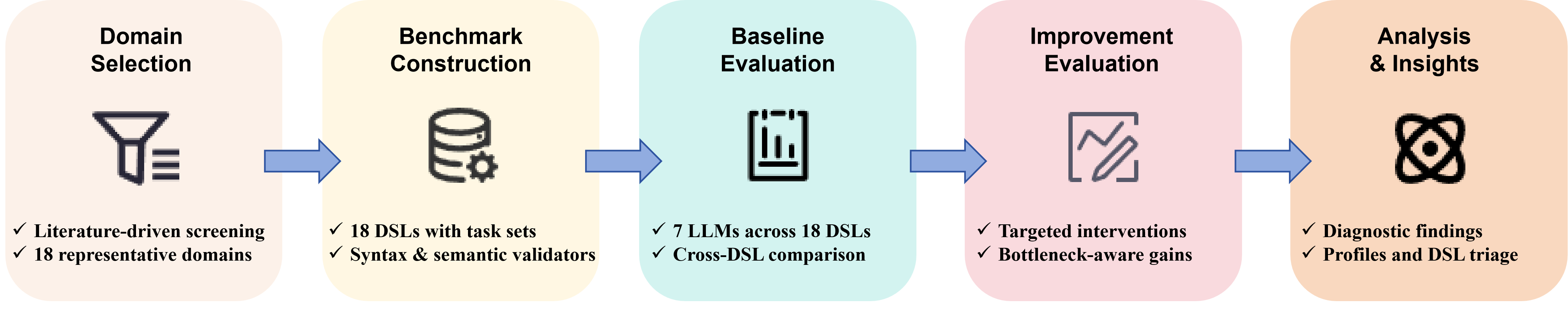}
\caption{DSL-Bench construction, evaluation, and analysis workflow.}
\label{fig:framework}
\end{figure*}

Figure~\ref{fig:framework} summarizes the overall DSL-Bench workflow and main outputs.

The results show that syntax acceptance substantially overestimates task-level correctness.
Across the 18 DSLs and seven evaluated LLMs, the mean syntax pass rate is 31.37\%, while the mean semantic pass rate is only 11.64\%.
LLM performance is also strongly shaped by the target DSL. Differences among DSLs account for 76.99\% of semantic pass-rate variation, compared with 7.86\% for differences among models.
Among the four DSL-level factors analyzed in this study, the Knowledge Demand factor has the strongest negative association with semantic correctness ($r=-0.508$).
Improvement gains are also bottleneck-dependent. Prompt engineering gives uneven gains across DSLs, while RAG-Refinement is the most effective targeted method in the five targeted DSLs.
These analyses further support diagnostic model capability profiles and a \textbf{DSL-Profiler} summary for data-scarce DSL triage.

In summary, this paper makes the following contributions:
\begin{itemize}
    \item We build DSL-Bench, an executable benchmark covering 18 representative DSLs, with syntax checkers and task-level semantic oracles for evaluating the same set of LLMs under the same testing conditions.
    \item We provide cross-DSL evidence on LLM code generation, showing a large gap between syntax acceptance and task-level semantic correctness and showing that performance varies more by DSL than by model.
    \item We analyze DSL-level factors and improvement methods, showing that Knowledge Demand is strongly associated with lower semantic correctness and that method gains are shaped by DSL-specific knowledge and semantic bottlenecks.
\item We derive diagnostic model capability profiles and DSL-Profiler summaries for triaging data-scarce DSLs.
\end{itemize}

The remainder of this paper is organized as follows.
Section~\ref{sec:background} gives the background needed for DSL-Bench.
Section~\ref{sec:rq1} describes the construction of DSL-Bench.
Section~\ref{sec:rq2} presents the baseline evaluation setup and results.
Section~\ref{sec:rq3} analyzes the observed performance variation, and Section~\ref{sec:rq4} evaluates improvement methods under DSL bottlenecks.
Section~\ref{sec:threats} discusses threats to validity, Section~\ref{sec:related} reviews related work, and Section~\ref{sec:discussion-conclusion} concludes.

\section{Background}
\label{sec:background}

\subsection{DSLs and Toolchains}

A DSL is designed around a particular application domain and usually trades general-purpose expressiveness for concise domain notation~\cite{mernik2005dsl,kosar2016dslmapping}.
DSLs appear as query languages, configuration languages, modeling languages, workflow languages, proof languages, hardware-description notations, and scientific-computing languages.
Their surface syntax may look familiar, but their meaning is defined by domain objects and by the toolchain that interprets them.
For example, a graph query is judged by query results, a proof script by proof checking, and a hardware description by simulation or synthesis behavior~\cite{feng2025cypherbench,yang2023leandojo,liu2023verilogeval}.

DSL ecosystems are often smaller and more fragmented than GPL ecosystems.
A DSL may have fewer public examples, multiple active versions, tool-specific extensions, or incomplete validation infrastructure.
The same surface notation may also behave differently across related languages or versions.
These properties make DSL generation different from generating GPL code with a domain library.
The generated program must match the domain notation, the domain semantics, and the constraints of the target toolchain.

\subsection{DSL Evaluation}

LLM-based code generation for low-resource and domain-specific programming languages has become an active research topic~\cite{joel2025survey}.
Existing DSL studies provide important evidence in individual settings, including hardware design, theorem proving, program verification, infrastructure configuration, smart contracts, and graph querying~\cite{liu2023verilogeval,yang2023leandojo,misu2024dafny,kon2024iaceval,peng2025soleval,feng2025cypherbench}.
These studies show that DSL evaluation needs executable or structural oracles, since a generated output may parse successfully while failing the domain task.

The current evidence is still fragmented across DSLs.
Existing benchmarks usually define tasks, validators, metrics, model sets, and execution environments for one language or one adjacent notation.
This makes it difficult to tell whether an observed failure comes from the model, the target DSL, the validator, or the evaluation setting.
A broader evaluation therefore needs results from multiple DSLs under the same model set and testing conditions, while still preserving the validator that each DSL requires.

\section{Benchmark Construction}
\label{sec:rq1}

A cross-DSL benchmark first needs a clear scope and a validation design that works across heterogeneous DSLs.
This leads to \textbf{RQ1. What evaluation scope is needed to obtain a broad and representative cross-DSL view of LLM code generation?}
The following subsections describe how DSL-Bench selects application domains, maps them to evaluable DSLs, constructs benchmark tasks, and builds syntax checkers and task-level semantic oracles.

\subsection{Domain Selection}

DSL-Bench starts from a literature-driven corpus assembled from arXiv, IEEE Xplore, Web of Science, the ACM Digital Library, and from 2022 to 2025 selected top-tier venues in software engineering, security, and artificial intelligence.
The pipeline screens DSL-related papers, applies LLM-assisted relevance filtering, deduplicates records, then uses open card sorting~\cite{begel2014analyze,nurmuliani2004cardsorting,hudson2014cardsorting,barrett1995knowledge,maiden1996acre,rugg1997sorting}, sentence-transformer clustering~\cite{reimers2019sentencebert}, and manual audit.
The screening process yields 79,844 candidates, 14,955 relevant records, 13,045 deduplicated papers, 4,796 labels, 15 pre-audit clusters, and 18 final domains covering 10,610 papers (81.3\%).
The 15-to-18 transition reflects split/merge decisions and multiple benchmarkable domains from coarse clusters. Four graduate students in software engineering participated in the manual review of the LLM-assisted clustering, domain labels, and domain-to-DSL mapping decisions. Exact queries, venues, prompts, sampled decisions, and mapping rationale are documented in the supplement.
The procedure is intended as reproducible selection, with taxonomy claims outside its scope.

\subsection{DSL Selection}
\label{sec:domain-mapping}

The literature screen identifies application domains, so we perform a criteria-guided manual mapping from each domain to a representative DSL.
Each final domain-to-DSL mapping is documented against the same five eligibility criteria: domain centrality, executable benchmarkability, task diversity, reproducibility with public resources, and non-redundancy.

This mapping yields 18 DSLs: Nextflow DSL2, SMILES, jq, Vega-Lite, Cypher, Terraform HCL, LaTeX, GDScript, Verilog, IEC 61131-3 Structured Text, SysML v2, Verilog-A, Dafny, OpenQASM 3, Behavior Tree XML, UFL, Solidity, and Lean.
The set mixes mature DSLs with newer or under-benchmarked ones and includes analytically useful pairs such as Verilog/Verilog-A.

\subsection{Dataset Construction}

For each DSL, we adapt an existing benchmark when available or construct tasks from public domain resources otherwise.
The resulting datasets fall into three categories.
The first category directly inherits existing benchmark data and evaluation methods, covering cheminformatics, data engineering, data visualization, database/query, DevOps/infrastructure, document processing, hardware design, program verification, quantum computing, robotics, smart contracts, and theorem proving~\cite{liu2023verilogeval,zheng2022minif2f,tsoukalas2024putnambench,peng2025soleval,feng2025cypherbench,luo2021nvbench,kale2025texpert,kon2024iaceval,izzo2024btgenbot,yang2025qcircuitbench,edwards2021text2mol,misu2024dafny,verbruggen2026jqbench}.
The second category uses existing data but rebuilds validators for this study, covering industrial automation and model-driven engineering.
The third category constructs both data and validators from public ecosystem resources, covering bioinformatics, game development, power electronics, and scientific computing~\cite{ditommaso2017nextflow,ewels2020nfcore,gdquest2023learngdscript,alnaes2014ufl,openvaf,liu2026agents4plc}.
The benchmark records source identifiers, sampling/adaptation status, validators, source task counts, and train/retrieval exclusions.

\DSLBenchOverviewTable

Table~\ref{tab:dslbench-overview} summarizes the benchmark coverage, source task pools, syntax validators, semantic validation bases, and construction categories.
For evaluation, each DSL is converted into an executable task set with a documented syntax validator and task-level semantic validation basis.
All reported pass rates are computed on the evaluated task sets used by the corresponding validators.
The supplement reports source-to-evaluation filtering, sampling, and adaptation details, together with per-DSL and per-model denominators.
Aggregate results are reported primarily as macro-averages over DSL-model cells to prevent large source pools from dominating the cross-DSL comparison, with evaluated-task-weighted aggregates reported as sensitivity checks.

\subsection{Validators}

DSL-Bench uses a two-level validator for every DSL.
Syntax pass rate means acceptance by the relevant DSL-specific parser, compiler, schema checker, proof assistant, simulator, or toolchain.
Semantic pass rate means satisfaction of a task-level semantic validation basis appropriate to the target DSL, such as execution, simulation, proof acceptance, canonical equivalence, query-result equivalence, policy checking, or structural control/dataflow checking.
DSL-Bench therefore unifies the generation protocol and syntax/semantic metric interface while preserving DSL-appropriate task oracles.
Cross-DSL comparison in DSL-Bench is diagnostic: it compares model behavior under a shared generation protocol and a common syntax/semantic metric interface, rather than assuming a single universal correctness notion across all DSLs.
Full commands, tool versions, audit notes, and validation boundaries are reported in the supplement.

To further check validator reliability, we manually audited 20 generated outputs for each DSL, stratified where possible across syntax failures, syntax-pass/semantic-fail cases, and semantic-pass cases.
For DSLs whose data and validators were newly constructed, we additionally reviewed all constructed benchmark items and validator rules during benchmark construction.
For each audited output, we inspected whether the validator decision matched the task intent and documented semantic validation basis.
The audit was used to check whether validator outcomes were consistent with the task specification rather than to override automated results.
For each audited case, reviewers inspected the prompt, generated output, validator output or log where available, and expected semantic criterion.
Disagreements or unclear cases are treated as validation-boundary notes rather than as manual overrides of the automated outcomes.

The project records separate benchmark construction inputs, generation outputs, validator outputs, and analysis scripts.
For each DSL, they record task identifiers, generation prompts, model outputs, syntax outcomes, semantic outcomes, and validator metadata.
This organization allows aggregate results to be traced back to per-task validator decisions and supports checks of the macro-average, evaluated-task-weighted sensitivity, factor analysis, and intervention results.

\findingbox{Answer to RQ 1}{A broad cross-DSL view of LLM code generation requires both an explicit DSL definition and executable validation; benchmark size by itself is insufficient. DSL-Bench instantiates this view with literature-driven coverage of 18 representative DSLs paired with a two-level syntax/semantic validator per DSL, so that parser acceptance and task-level semantic correctness become separately observable and aggregable under a common protocol.}

\section{Baseline Evaluation}
\label{sec:rq2}

With DSL-Bench constructed, the next step is to measure LLM performance across the selected DSLs.
This leads to \textbf{RQ2. How does LLM performance vary across DSLs when syntax acceptance and task-level semantic correctness are measured under the same models and testing conditions?}
This section reports the model setup, evaluation metrics, baseline pass rates, DSL/model effects, model ranking variation, and model capability profiles.

\subsection{Model Setup}

We evaluate seven LLMs that cover different model families, sizes, and deployment settings: Qwen3-1.7B, Qwen3-8B, Qwen3-32B, Qwen2.5-Coder-7B, Ministral-3-8B-Instruct, DeepSeek-v3.2, and GLM-4.7.
This set includes a controlled Qwen size progression, a code-specialized model, compact open models, and larger general-purpose models.
Local inference for the locally deployable models---Qwen3-1.7B, Qwen3-8B, Qwen3-32B, Qwen2.5-Coder-7B, and Ministral-3-8B-Instruct---was performed on an NVIDIA DGX Spark environment with 128~GB GPU memory.
DeepSeek-v3.2 and GLM-4.7 were evaluated through online API endpoints because they were not locally deployable in our experimental setup.
For all models, we used the same prompt templates, output extraction rules, validation pipeline, and metric computation; for remotely served models, we recorded the endpoint model name, query date, and decoding parameters where available.
We focus on reproducible and deployable model configurations rather than closed-source frontier upper bounds. Evaluating frontier APIs across all 18 DSLs and intervention settings would substantially increase cost and introduce versioning and reproducibility issues, so systematic frontier-model evaluation is outside the scope of this study.
All baseline generations use zero-shot prompting with temperature 0 and deterministic decoding where supported.

\subsection{Evaluation Metrics}

For each DSL \(d\), model \(m\), and task \(t\), the validator returns two binary outcomes, \(SYN_{d,m,t}\) and \(SEM_{d,m,t}\).
\(SYN_{d,m,t}\) indicates whether the generated code passes syntax validation, and \(SEM_{d,m,t}\) indicates whether it passes semantic validation.
The syntax pass rate for a DSL-model cell is the percentage of tasks accepted by the DSL parser, compiler, schema checker, proof assistant, or equivalent toolchain.
The semantic pass rate is the percentage of tasks satisfying the DSL-specific semantic oracle.
We also report the syntax-semantic gap, defined as syntax pass rate minus semantic pass rate.
Unless otherwise stated, aggregate results are macro-averaged over DSL-model cells so that DSLs with larger evaluated task sets, such as Solidity, Lean, Cypher, or SMILES, have the same cell-level weight as smaller DSLs of practical interest, such as UFL, Verilog-A, or GDScript.
The supplementary material reports the corresponding evaluated-task-weighted aggregate as a sensitivity check.

\subsection{Overall Results}

\begin{figure*}[t]
\centering
\includegraphics[width=\textwidth]{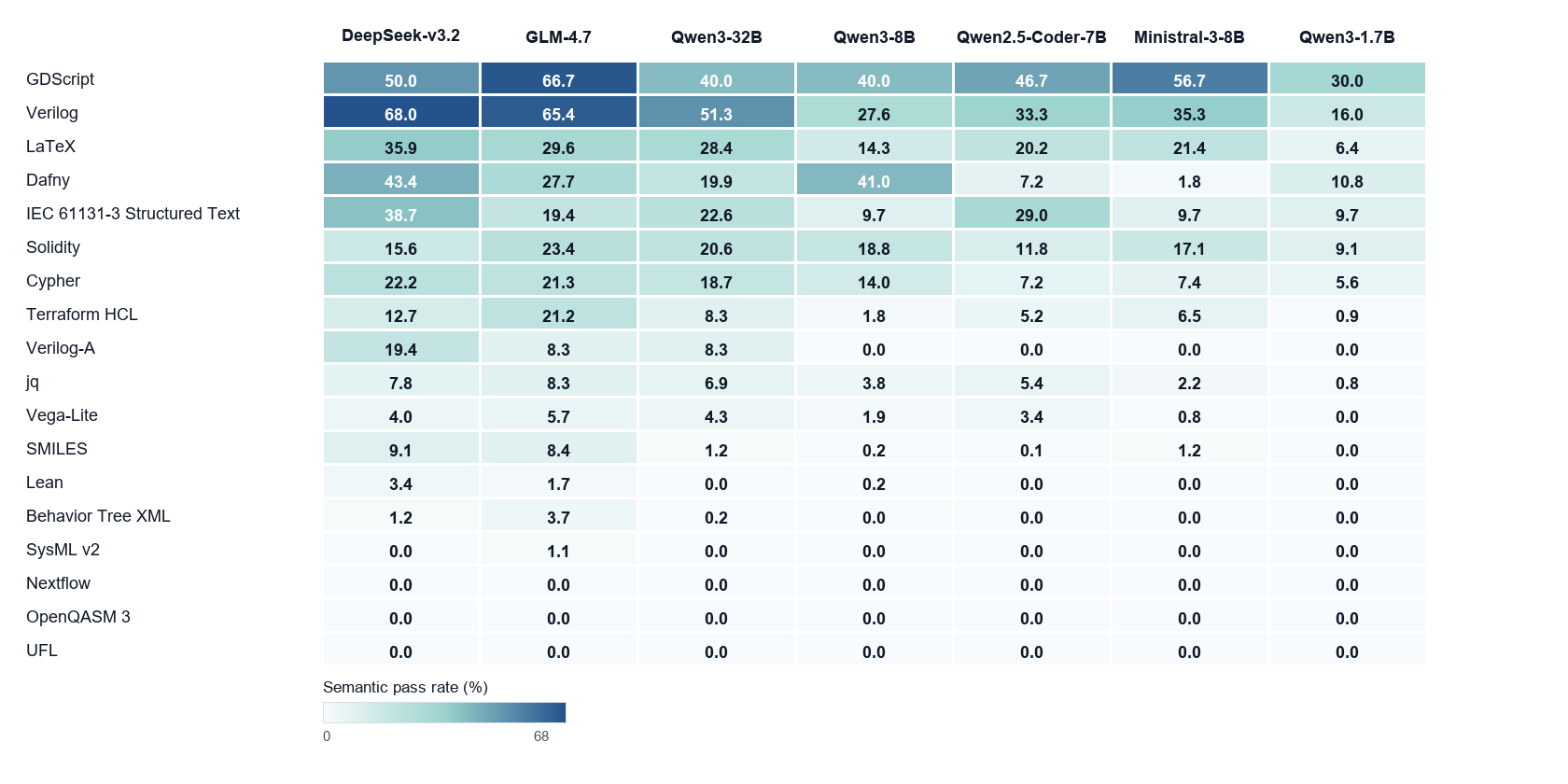}
\caption{Baseline semantic pass rates across DSLs and evaluated models.}
\label{fig:semantic-heatmap}
\end{figure*}

Figure~\ref{fig:semantic-heatmap} shows the baseline semantic pass rates by DSL and deployment configuration.
At the model level, DeepSeek-v3.2 and GLM-4.7 have the highest macro semantic pass rates, while the full model-level baseline table is provided in the supplement.
Across all 126 DSL-model cells, the mean syntax pass rate is 31.37\%, while the mean semantic pass rate is only 11.64\%.
The mean syntax-semantic gap is 19.74 percentage points.
This gap shows that a substantial share of generated DSL programs are accepted by the relevant toolchain and still fail the task-level oracle.
The result supports the central motivation for DSL-Bench.
Syntax-only or compilation-only evaluation can overestimate DSL code generation competence by a factor of about three on average.
Under evaluated-task-weighted aggregation, reported in the supplementary material, the corresponding baseline mean is 33.13\% syntax and 8.10\% semantic, preserving the same qualitative syntax-semantic gap.

\subsection{Per-DSL Results}

Performance varies far more across DSLs than the overall average suggests.
The strongest DSLs are GDScript and Verilog, with mean semantic pass rates of 47.14\% and 42.40\%, respectively.
In contrast, Bioinformatics/Nextflow, Scientific Computing/UFL, and Quantum Computing/OpenQASM 3 have zero mean semantic pass rate under the baseline setting, while SysML v2 and Behavior Tree XML are near zero.

The DSL-level results reveal two patterns.
First, a low semantic score can arise from very different failure modes.
OpenQASM 3 and UFL fail almost entirely at the syntax/toolchain layer, whereas Nextflow and Behavior Tree XML often have nontrivial syntax pass rates alongside near-zero semantic correctness.
Second, surface familiarity gives limited guidance about semantic success.
DSLs such as Vega-Lite, SMILES, Cypher, Nextflow, and Behavior Tree XML often produce large syntax-semantic gaps, suggesting that the model can generate plausible surface forms without satisfying the domain-level intent.
Several DSLs have small evaluated task sets, including UFL (25), GDScript (30), IEC 61131-3 ST (31), and Verilog-A (36), matching their source task pools in Table~\ref{tab:dslbench-overview}.
Their per-DSL rates should therefore be read as high-uncertainty estimates; a single task changes the semantic pass rate by 2.8 to 4.0 percentage points in these DSLs.
We keep them in the macro analysis because they cover under-benchmarked settings of practical interest, and we separately report the leave-small-DSL-out sensitivity analysis below.

\subsection{DSL and Model Effects}

To quantify whether performance variation is associated more with the target DSL or with the model, we decompose DSL-model cell performance using a balanced two-way additive ANOVA over the 18 DSLs and seven models.
For each DSL-model cell \(y_{d,m}\), where \(y\) is either syntax or semantic pass rate, we fit
\[
y_{d,m}=\mu+\alpha_d+\beta_m+\epsilon_{d,m}.
\]
Here \(\alpha_d\) is the DSL effect, \(\beta_m\) is the model effect, and the residual captures DSL-model interaction and unexplained variation.
We report eta-squared proportions for DSL, model, and residual components.
We estimate uncertainty by bootstrap resampling DSL-model cells.
Because DSL source task pools range from 25 to 33,008 tasks and evaluated task counts also remain imbalanced after sampling/adaptation, this ANOVA provides a macro cell-level comparison over DSL-model conditions.
Evaluated task counts and evaluated-task-weighted sensitivity checks are reported in the supplement.
Table~\ref{tab:variance} reports the resulting eta-squared proportions.
For semantic correctness, the DSL main effect accounts for 76.99\% of macro-cell pass-rate variation, while the model main effect accounts for 7.86\%.
For syntax pass rate, the DSL main effect still dominates, accounting for 64.53\% of variation, while the model main effect accounts for 15.94\%.
Bootstrap confidence intervals and evaluated-task-weighted sensitivity checks show the same ordering.
Under evaluated-task weighting, the DSL main effect accounts for 72.49\% of semantic variation, compared with 11.80\% for the model main effect.
This additive view isolates the relative magnitude of DSL and model main effects, while ranking instability and Kendall's \(W\) still show DSL-by-model interaction.

\begin{table}[t]
\centering
\caption{Additive DSL/model main-effect decomposition with bootstrap 95\% confidence intervals.}
\label{tab:variance}
\scriptsize
\begin{tabularx}{\linewidth}{lX X r}
\toprule
Metric & DSL \(\eta^2\) & Model \(\eta^2\) & Resid. \\
\midrule
Semantic & 76.99 [71.46, 87.27] & 7.86 [4.23, 23.35] & 15.15 \\
Syntax & 64.53 [58.11, 79.37] & 15.94 [9.45, 32.26] & 19.53 \\
\bottomrule
\end{tabularx}
\end{table}

To check whether the conclusion is driven by small DSLs or validation-scope differences, we rerun the same cell-level decomposition under two restrictions.
Removing DSLs with fewer than 50 evaluated tasks leaves 14 DSLs and preserves the same ordering: the DSL main effect accounts for 71.91\% of semantic variation and the model main effect accounts for 9.57\%.
Restricting to the 11 DSLs with fully executable task-level semantic validation gives a similar result. The DSL main effect accounts for 77.51\% and the model main effect accounts for 8.17\%.

To check that the macro-cell ANOVA does not drive this conclusion, we also fit a binomial mixed-effects sensitivity model over DSL-model semantic pass counts.
The estimated DSL random-effect variance is 15.03 times larger than the model random-effect variance, preserving the same qualitative ordering.
Full model details are reported in the supplement.

Model choice remains relevant.
Larger or stronger models generally perform better on average.
The result shows that model improvements are strongly conditioned by the target DSL.
Moving from one DSL to another changes semantic correctness more than moving from one evaluated model to another in this benchmark.

\subsection{Model Rankings}

If model capability transferred uniformly across DSLs, the model ranking would be stable from one DSL to another.
Instead, rankings shift from DSL to DSL.
Kendall's coefficient of concordance over per-DSL semantic rankings is \(W=0.426\), indicating limited agreement across DSLs.
The strongest average models, DeepSeek-v3.2 and GLM-4.7, lead on multiple DSLs, and smaller or code-specialized models remain competitive in particular settings.
This reinforces the variance-decomposition result: model choice matters, and its effect depends on the DSL.

\subsection{Capability Profiles}

We use the DSL--model matrix to derive exploratory capability profiles.
These profiles normalize model performance within DSLs and summarize relative strengths across bottleneck categories.
They are diagnostic summaries rather than calibrated model rankings; small differences are interpreted with bootstrap confidence intervals and tied-within-CI comparisons.
The full formulas, axis definitions, scorecard figure, and results are in the supplement.

\findingbox{Answer to RQ 2}{Across 18 DSLs and 7 LLMs, the mean semantic pass rate is 11.64\% with a 19.74-point syntax/semantic gap, and the DSL main effect explains 76.99\% of macro-cell semantic variation versus 7.86\% for the model effect; Kendall's $W=0.426$ further reveals substantial DSL-by-model interaction. Exploratory bottleneck-axis profiles can supplement, but not replace, per-DSL results.}

\section{Variation Analysis}
\label{sec:rq3}

The baseline results show that the target DSL explains much of the semantic performance variation.
This leads to \textbf{RQ3. Which DSL-level properties and failure modes help explain cross-DSL performance variation?}
This section analyzes DSL-level factors, presents DSL-Profiler, and examines qualitative failure cases.

\subsection{DSL Factors}

We compute four normalized DSL-level proxies from benchmark/reference corpora and metadata.
KD captures domain terminology, local API/symbol surface, and domain-term concentration.
SF captures surface similarity to common programming, query, configuration, and markup notations.
HC captures executable scaffolding and environment burden.
Release Year approximates ecosystem recency and version instability.
The supplement gives the extraction formulas and raw component values.
The factor analysis relates these DSL-level proxies to DSL-level mean syntax and semantic pass rates.
We report Pearson and Spearman correlations.
Because there are 18 DSLs, we treat these correlations as explanatory evidence and avoid causal claims.
We check stability with bootstrap, permutation, and leave-one-DSL-out analyses.

\FactorSummaryTable

Table~\ref{tab:factor-correlations} summarizes the fitted bivariate slopes and correlations.
Knowledge Demand has the largest observed negative association with semantic correctness among the four corpus-derived factors.
This association is directionally stable under leave-one-DSL-out, partial-correlation, bootstrap, and permutation checks.
The bootstrap interval for Pearson \(r\) is \([-0.770,-0.006]\), and permutation testing gives \(p=0.030\), but this uncorrected \(p\)-value fails a conservative Bonferroni correction.
We therefore treat KD as the strongest explanatory factor while avoiding definitive or causal claims.
Harness Complexity is more strongly associated with syntax/toolchain validity than with semantic correctness.
Syntax Familiarity is weakly associated with semantic correctness, suggesting that surface resemblance to common code gives limited guidance.
Release year is negatively correlated with performance, consistent with newer DSLs having less public ecosystem exposure and stronger version instability.

\subsection{DSL-Profiler}

Practical DSLs often lack public task sets, complete validators, or enough examples for a full DSL-Bench-style evaluation.
DSL-Profiler is intended as a diagnostic triage aid rather than a calibrated pass-rate predictor.
It maps a target DSL's standardized factor profile to nearby benchmark DSLs and returns coarse risk ranges, a bottleneck hypothesis, a candidate intervention, and nearest-neighbor distance as a trust indicator.
Leave-one-DSL-out validation gives a semantic MAE of 11.78 percentage points and a syntax MAE of 20.29 points, so the output should be read as a triage hint that guides which validators, retrieval sources, or repair loops to inspect first.
The formulas and nearest-neighbor details are in the supplement.

\subsection{Knowledge Bottlenecks}

High-KD DSLs tend to require domain concepts that are difficult to infer from surface syntax.
UFL requires finite-element and variational-form knowledge; Verilog-A requires analog circuit modeling and continuous-time physical semantics; Vega-Lite requires visualization intent and field/channel semantics; and Cypher requires graph-schema and query-result semantics.
These DSLs require more than familiarity with token sequences.
Correct programs must bind domain entities, APIs, or mathematical objects in ways that satisfy an executable or structural oracle.

The correlation between Knowledge Demand and semantic correctness is therefore consistent with the main thesis: observed DSL failures often reflect weak semantic grounding.
We interpret KD as an explanatory signal and make no causal claim from it.
It captures observable properties of the benchmark/reference corpora, and it may interact with task design, dataset size, public documentation, and model pretraining exposure.
We therefore use KD together with case analyses and avoid treating it as a single universal difficulty score.

\subsection{Misleading Syntax}

Syntax Familiarity alone is a weak predictor of semantic correctness.
This leads to a recurring phenomenon that we call \emph{syntactic deception}: a DSL appears transferable because its surface form resembles a familiar language, while semantic correctness collapses because the DSL requires different domain-specific execution semantics.

Verilog and Verilog-A illustrate this effect.
Both languages share Verilog-family lexical and module-level conventions, such as module declarations, port lists, parameter declarations, and \texttt{endmodule}-delimited structure.
Verilog-A additionally introduces analog blocks, electrical disciplines, branch declarations, and contribution statements whose correctness depends on continuous-time physical semantics.
In DSL-Bench, Verilog has one of the highest baseline semantic pass rates, while Verilog-A remains much harder.
Surface resemblance alone leaves the difference unexplained.
It reflects a shift from digital RTL behavior, where simulation testbenches can often validate discrete logic, to analog behavioral modeling, where physical relationships, disciplines, and simulator-compatible contributions must be correct.

Syntactic deception matters for AI coding tools because a plausible-looking DSL output can be actively misleading.
The code may contain familiar declarations and delimiters while violating the domain semantics that the target toolchain or simulator enforces.
This helps explain why the syntax pass rate often runs two to three times the semantic pass rate within a DSL.

\subsection{Structural Semantics}

Knowledge Demand explains only part of the failure pattern.
Behavior Tree XML exposes a complementary structural-semantics bottleneck.
It has relatively low Knowledge Demand, low Harness Complexity, and high Syntax Familiarity, yet its baseline semantic pass rate is only 0.73\%.
The reason is that correctness depends on the composition of the tree, beyond producing parseable XML.

Behavior-tree tasks require the model to choose and place control-flow nodes, conditions, actions, decorators, blackboard keys, and functional attributes so that the resulting tree matches the intended robot behavior.
An output can be valid XML and accepted by BehaviorTree.CPP while still failing because the selector/sequence topology is wrong, a functional node is missing, or blackboard dataflow does not match the reference behavior.
This case shows that DSL failures can arise even without dense domain terminology.
Some arise from structural semantics: the meaning of the program is encoded in topology, ordering, dependencies, and role-specific node composition.

This observation has direct implications for future DSL agents.
Retrieval alone can make outputs look more like reference examples, while structural DSLs likely need topology-aware planning and validator-guided repair.
For such DSLs, the central challenge is assembling the right executable structure alongside retrieving the right vocabulary.

\subsection{Version Interference}

Some DSLs also fail because model priors are pulled toward older or adjacent language versions.
SysML v2 is a clear example.
It is newer than SysML v1 and uses a different textual modeling syntax; LLMs often generate UML-like or SysML v1-style fragments.
OpenQASM 3 similarly differs from earlier QASM conventions, and generated programs may mix old-version syntax with newer task requirements.

Legacy version interference is distinct from low-resource scarcity.
A model may have substantial exposure to an older language version or adjacent ecosystem, making the wrong prior stronger than the correct one.
This explains why explicit version constraints, retrieved local examples, or fine-tuning can help in some cases, while generic prompting may have limited ability to override parameterized prior knowledge.

We therefore treat legacy version interference as a qualitative failure mode rather than estimating it as a separate quantitative factor.
The supplemental material includes representative evaluator-log examples for these failure modes, including Verilog-A syntactic deception, Behavior Tree XML structural mismatch, SysML v2 legacy-version interference, and UFL local-API failures.

\findingbox{Answer to RQ 3}{Among four DSL-level proxies, Knowledge Demand shows the largest negative association with semantic correctness (Pearson $r=-0.508$), and case analyses surface three recurring failure modes: syntactic deception, structural-semantics bottlenecks, and legacy-version interference. DSL-Profiler uses these exploratory profiles to provide bottleneck hypotheses and candidate-intervention triage hints for unseen DSLs.}

\section{Improvement Evaluation}
\label{sec:rq4}

The preceding analyses show that DSL performance depends on DSL-specific knowledge, toolchains, and semantic bottlenecks.
This leads to \textbf{RQ4. How do common LLM improvement methods behave under different DSL bottlenecks?}
This section evaluates prompt engineering across all 18 DSLs and targeted retrieval, refinement, and adaptation methods on bottleneck-selected DSLs.

\subsection{Improvement Methods}

We evaluate three intervention spaces.
First, benchmark-wide prompt engineering (PE) is applied to all 18 DSLs using DSL-specific instructions, output constraints, and non-benchmark examples.
Second, targeted non-finetuning pipelines are compared on five bottleneck-selected DSLs.
RAG retrieves local DSL documentation or examples before generation.
Baseline-Refinement repairs baseline output using evaluator feedback.
Retrieval-Refinement adds retrieved context to repair.
RAG-Refinement repairs a retrieval-generated output using both retrieved context and evaluator feedback.
Third, a separate Qwen-family LoRA adaptation study is reported in the supplement.

The five deep-study DSLs are SysML v2, Verilog-A, UFL, Behavior Tree XML, and Terraform HCL.
They cover version interference, high-KD/high-HC domain semantics, structural semantics, and configuration-policy bottlenecks.
Because these interventions require retrieval corpora, evaluator feedback traces, or training data, the five-DSL study provides bottleneck-aware evidence without claiming benchmark-wide method superiority.
The supplemental bottleneck-profile table gives the detailed selection rationale.

For intervention experiments, we compute gains relative to the matching baseline for the same DSL and model.
For method \(h\), the semantic gain is
\[
\Delta^{sem}_{h}(d,m)=
Sem_{h}(d,m)-Sem_{baseline}(d,m).
\]
Syntax gains are computed analogously.
This paired DSL-model comparison controls for the wide variation in baseline difficulty across DSLs and models.
Where task-level paired evaluator outputs are available, we additionally use paired bootstrap confidence intervals and exact McNemar tests.
The supplementary material reports these paired tests for the fully paired targeted intervention subset and separately reports model-domain averages for the two extension DSLs.

\subsection{Improvement Results}

\begin{figure}[!b]
\centering
\includegraphics[width=\linewidth]{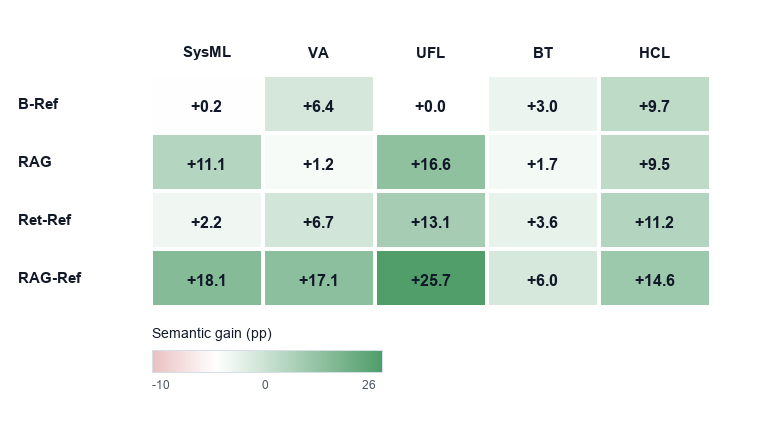}
\caption{Semantic gains of targeted interventions over matched baselines; B-Ref and Ret-Ref denote Baseline-Refinement and Retrieval-Refinement.}
\label{fig:intervention-heatmap}
\end{figure}

Figure~\ref{fig:intervention-heatmap} summarizes the five-DSL semantic gains.
Across all 126 DSL-model rows, benchmark-wide PE improves mean syntax pass rate by 6.38 percentage points and semantic pass rate by 2.42 percentage points over the matching baseline.
This confirms that careful prompting is a useful baseline intervention, although the average hides strong heterogeneity.
Semantic gains are positive in 12 of 18 DSLs and negative in 6, with 65 model-domain rows improving, 23 unchanged, and 38 regressing.
We therefore treat PE as a benchmark-wide output-discipline baseline whose benefit varies by DSL.
The full per-DSL PE gains are in the supplement; the negative cases suggest that generic persona, internal reasoning instructions, and few-shot demonstrations can conflict with strict output formats, validator expectations, or strong model priors.

In the five targeted DSLs, RAG-Refinement improves semantic correctness by 16.29 percentage points on average and syntax by 34.34 percentage points.
Because the deep-study DSLs include low-baseline bottleneck settings, this gain should be read within the selected setting.
Direct RAG is second in semantic gain (+8.02 points), followed by Retrieval-Refinement (+7.37) and Baseline-Refinement (+3.83).
The supplementary material gives the full five-DSL method table.
For the three DSLs with task-level paired evaluator JSONs (SysML v2, UFL, and Verilog-A), RAG-Refinement improves semantic correctness over the matching baseline by 19.11 points with a paired-bootstrap 95\% CI of [16.65, 21.57] and exact McNemar \(p<10^{-50}\).
For Behavior Tree XML and Terraform HCL, the evaluator outputs are aggregated at model-domain granularity, so we report model-domain averages rather than task-level McNemar tests.

We report PE separately from retrieval/refinement methods because PE is evaluated over all 18 DSLs, whereas retrieval and refinement require additional DSL-specific infrastructure and are evaluated in the five deep-study DSLs. The Qwen-family LoRA study is separate from both comparisons.

Within this five-DSL setting, RAG-Refinement remains the top method after holding out any one intervention DSL, although dominance outside this set remains untested.
The method combines retrieval for DSL vocabulary, APIs, versions, and examples with evaluator-guided repair of task-specific failures.
DSL-Profiler converts these intervention results into bottleneck-aware triage hints.

\subsection{Retrieval Limits}

Direct RAG improves syntax validity, while semantic gains are smaller because retrieval injects knowledge without guaranteeing task alignment.
In Behavior Tree XML, retrieved examples make outputs look like valid BehaviorTree.CPP XML without ensuring correct control-flow topology or blackboard dataflow; in Terraform HCL, snippets can improve resource syntax while missing policy intent.
RAG-Refinement performs better because retrieval moves the model toward the right DSL constructs and evaluator feedback then repairs task-specific failures.

\subsection{Gain Patterns}

Feedback-only refinement is weakest on average because it repairs local errors without supplying missing DSL concepts.
Model capacity also modulates gains.
RAG-Refinement has the strongest scale association (\(r=0.761\) with \(\log_{10}\) parameter count).
A separate Qwen-family LoRA adaptation study is reported in the supplement.

\findingbox{Answer to RQ 4}{Intervention gains depend on the DSL bottleneck. Benchmark-wide prompt engineering improves the mean unevenly (semantic $+2.42$ pp, with 12/18 DSLs gaining and 6 regressing). In the five targeted DSLs, RAG-Refinement improves semantic correctness by 16.29 percentage points on average. These targeted results do not establish benchmark-wide method superiority.}

\section{Threats to Validity}
\label{sec:threats}

\textbf{Construct validity.}
Our broad DSL definition leaves domain-to-DSL mappings contestable, and the $n{=}18$ factor analysis, capability profiles, and DSL-Profiler are explanatory and relative, without calibrated predictions.

\textbf{Benchmark construction and contamination.}
LLM-assisted screening and card sorting may miss domains, and public benchmark-derived tasks may appear in pretraining; we exclude target solutions from retrieval and adaptation and run a local overlap screen, while residual contamination remains possible.

\textbf{Oracle heterogeneity.}
Validators differ across DSL ecosystems, so semantic pass rates are documented oracle outcomes within each DSL; identical difficulty scoring across DSLs is outside the claim.

\textbf{Model coverage and deployment.}
The evaluated models do not include all closed-source frontier coding models. This is a deliberate scope choice motivated by reproducibility, deployment control, version stability, and the cost of evaluating many DSL-task-intervention combinations. Local inference used an NVIDIA DGX Spark environment, while DeepSeek-v3.2 and GLM-4.7 used online API endpoints, so the results characterize the selected deployment configurations rather than the absolute upper bound of current LLM coding capability.

\textbf{Interventions and profiler.}
Retrieval and refinement are evaluated on five targeted DSLs, while prompt engineering covers all 18 DSLs. DSL-Profiler offers exploratory risk ranges and triage hints without guaranteeing improvement; sensitivity checks, paired tests, and profiler validation are in the supplement.

\section{Related Work}
\label{sec:related}

\subsection{GPL Benchmarks}

GPL code generation benchmarks vary task scale, language coverage, project context, contamination control, and test-oracle design~\cite{chen2021codex,austin2021mbpp,hendrycks2021apps,li2022codecontests,lu2021codexglue,cassano2023multiple,lai2023ds1000,jimenez2024swebench,yu2024codereval,du2024classeval,liu2023evalplus,jain2025livecodebench,zhuo2025bigcodebench}.
Domain-oriented benchmarks further show that code generation ability is domain-conditioned~\cite{zhu2025domaineval,zheng2026domaincodebench,gu2025domcoder}.
These benchmarks mainly evaluate GPL code or GPL code that uses domain libraries.
Their results show that LLM coding performance depends on task form, context, tests, and execution environment.

\subsection{Executable Oracles}

Executable benchmark infrastructure studies how benchmark tasks should be paired with reliable checking mechanisms.
Defects4J and BugsInPy show the importance of reusable benchmark tooling~\cite{just2014defects4j,widyasari2020bugsinpy}, while EvalPlus and LiveCodeBench show that weak tests and contamination can distort LLM-code conclusions~\cite{liu2023evalplus,jain2025livecodebench}.
These works establish oracle quality, task reproducibility, and contamination control as central concerns in code generation evaluation.

\subsection{DSL Benchmarks}

The DSL literature characterizes DSLs as notations aligned with domain abstractions and tool-support assumptions~\cite{mernik2005dsl,kosar2016dslmapping}.
This literature shows that DSLs differ from GPLs not only in syntax, but also in domain concepts, toolchains, and semantic interpretation.
A recent survey by Joel et al. reviews LLM-based code generation for low-resource and domain-specific programming languages, covering datasets, models, evaluation practices, and adaptation methods~\cite{joel2025survey}.
Prior LLM studies evaluate individual DSLs or adjacent executable notations in hardware and verification, theorem proving, infrastructure as code, smart contracts, quantum and system modeling, constrained SQL, PLC Structured Text, and related settings~\cite{delgado2026constraintdsl,liu2023verilogeval,wu2025rtlrepocoder,zheng2022minif2f,tsoukalas2024putnambench,yang2023leandojo,misu2024dafny,kon2024iaceval,jana2026terraformer,peng2025soleval,ye2025solcontracteval,guo2025quanbench,jin2025sysmbench,scholak2021picard,fakih2024llm4plc,haag2024stfeedback,liu2026agents4plc}.
Related resources cover query, visualization, document, robotics, quantum, molecule, workflow, scientific-computing, game-scripting, analog-modeling, and JSON-transformation settings~\cite{feng2025cypherbench,luo2021nvbench,kale2025texpert,izzo2024btgenbot,yang2025qcircuitbench,edwards2021text2mol,ditommaso2017nextflow,ewels2020nfcore,alnaes2014ufl,gdquest2023learngdscript,openvaf,verbruggen2026jqbench}.
These studies establish the value of specialized evaluation, usually with tasks, validators, metrics, and model sets designed around one DSL or one adjacent notation.

However, this specialization also limits the questions that can be answered by single-DSL studies alone.
When each benchmark uses its own task distribution, validator design, model set, and reporting convention, it becomes difficult to determine whether an observed result is driven by the model, the target DSL, the toolchain, or the oracle.
DSL-Bench complements these studies by preserving DSL-specific validators while normalizing the generation protocol and the syntax/semantic metric interface across a broader set of target notations.
Its purpose is therefore not to replace domain-specific benchmarks, but to make their results comparable along common axes of variation.

\subsection{Improvement Methods}

Prompting, constrained decoding, retrieval, validator-feedback refinement, self-debugging, and LoRA-style adaptation target different failure modes~\cite{lewis2020rag,chen2023codet,chen2024selfdebug,madaan2023selfrefine,shinn2023reflexion,yao2023react,hu2022lora}.
Low-resource-language work likewise suggests that fine-tuning, in-context learning, and translation-style objectives operate in different regimes~\cite{giagnorio2025lrl}.
These methods are often evaluated within one language, model family, or benchmark setting, so reported gains depend on the target environment and available feedback or training data.

This dependence on the target setting is especially important for DSLs.
A method that improves surface form may not improve task semantics if the main bottleneck is missing domain knowledge, version mismatch, or toolchain setup.
Conversely, retrieval or feedback may help only when the retrieved material and validator signal are aligned with the DSL version, execution environment, and expected semantic behavior.
For this reason, DSL-Bench treats improvement methods as bottleneck-sensitive interventions rather than as benchmark-independent improvements.

\section{Discussion and Conclusion}
\label{sec:discussion-conclusion}

DSL-Bench suggests a practical design lesson for DSL-aware coding agents: surface validity leaves important errors undetected.
The gap between syntax acceptance and semantic correctness shows that parser acceptance must be paired with task oracles such as proofs, simulations, query answers, policy checks, or structural validators.
The variance decomposition also indicates that DSL bottlenecks are a useful organizing principle for support: high-KD DSLs need documentation grounding, syntactically deceptive DSLs need version- and semantics-aware constraints, and structural DSLs need topology and dataflow checks.

This observation is also relevant to the current shift from general coding assistants toward domain-specific agents.
For such agents, improving the base model alone is unlikely to be sufficient.
DSL-specific support needs to cover at least three sources of failure exposed by DSL-Bench.
The first is syntactic and version-specific knowledge, including grammars, target-version constructs, output templates, and invalid surface patterns that can be constrained before or during generation.
The second is domain knowledge, including concepts, APIs, invariants, design idioms, and task semantics that cannot be recovered from syntax alone.
The third is the runtime and toolchain environment, including parsers, compilers, simulators, proof assistants, policy checkers, and diagnostic logs that make validator-grounded repair possible.

These distinctions give a concrete basis for designing domain-specific agents.
Rather than treating all DSL tasks as ordinary code generation, an agent can select retrieval sources, prompting strategies, repair actions, and validation tools according to the bottleneck profile of the target DSL.
Knowledge-heavy DSLs call for documentation- and example-grounded generation; structurally constrained DSLs require explicit checks over topology, dependency, and dataflow; and toolchain-sensitive DSLs require environment-aware execution and version control.
DSL-Bench supports this line of work by reporting syntax and semantic outcomes separately and by exposing where surface validity fails to imply task correctness.

The improvement results reinforce this view.
Prompting improves output discipline, retrieval supplies missing concepts, feedback repairs executable failures, and their combination is most useful when the model can operationalize retrieved context.
Future systems should combine retrieval, executable validation, and iterative repair while reporting syntax and semantic metrics separately.

In conclusion, DSL-Bench provides a literature-driven executable benchmark across 18 DSLs and shows that DSL code generation is largely a semantic-grounding problem in our benchmark.
Progress will require DSL-specific corpora, validators, and agent workflows that reason over domain semantics in addition to syntax.

\bibliographystyle{IEEEtran}
\bibliography{reference}

\end{document}